\documentclass[aps, prb, twocolumn, superscriptaddress, showpacs, showpacs,preprintnumbers,amsmath,amssymb]{revtex4-2}
\usepackage{amssymb}
\usepackage{amsmath}
\usepackage{bm}
\usepackage[dvips]{graphicx}
\usepackage{dcolumn}
\usepackage{longtable}
\usepackage{color}

\begin{document}

\title{Pairing interaction in superconducting UCoGe tunable by magnetic field}
\author{K.~Ishida}
\email{kishida@scphys.kyoto-u.ac.jp}
\author{S.~Matsuzaki}
\author{M.~Manago}
\author{T.~Hattori}
\author{S.~Kitagawa}
\affiliation{Department of Physics, Graduate School of Science, Kyoto University, Kyoto 606-8502, Japan}
\author{M.~Hirata}
\author{T.~Sasaki}
\affiliation{Institute for Materials Research, Tohoku University. Sendai 980-8577, Japan}
\author{D.~Aoki}
\affiliation{Institute for Materials Research, Tohoku University, Oarai, Ibaraki 311-1313, Japan}
\affiliation{Universit\'e Grenoble Alpes, CEA, IRIG, PHELIQS, F-38000 Grenoble, France}
\date{\today}
\begin{abstract}
The mechanism of unconventional superconductivity, such as high-temperature-cuprate, Fe-based, and heavy-fermion superconductors, has been studied as a central issue in condensed-matter physics. 
Spin fluctuations, instead of phonons, are considered to be responsible for the formation of Cooper pairs, and many efforts have been made to confirm this mechanism experimentally. 
Although a qualitative consensus seems to have been obtained, experimental confirmation has not yet been achieved. 
This is owing to a lack of the quantitative comparison between theory and experiments.
Here, we show a semi-quantitative comparison between the superconducting-transition temperature ($T_{\rm SC}$) and spin fluctuations derived from the nuclear magnetic resonance (NMR) experiment on the ferromagnetic (FM) superconductor UCoGe, in which the FM fluctuations and superconductivity are tunable by external fields.
The enhancement and abrupt suppression of $T_{\rm SC}$ by applied fields, as well as the pressure variation of $T_{\rm SC}$ around the FM criticality are well understood by the change in the FM fluctuations on the basis of the single-band spin-triplet theoretical formalism.
The present comparisons strongly support the theoretical formalism of spin-fluctuation-mediated superconductivity, particularly in UCoGe. 
\end{abstract}

\pacs{74}
\maketitle
\section{Introduction}
U-based ferromagnetic (FM) superconductors\cite{SaxenaNature2000, AokiJPSJ2019Rev} and the recently discovered UTe$_2$\cite{RanScience2019} similar to them have attracted much attention and have been studied intensively.
This is because superconductivity becomes stronger under external fields\cite{AokiJPSJ2014Rev}, which is not explained by spin-singlet pairing, and spin-triplet superconductivity is expected. 
The identification of the spin-triplet superconducting (SC) state and unveiling of the SC pairing mechanism of unconventional superconductivity, which have been considered to be other than the electron-phonon interaction, are central issues in condensed-matter physics.   
From these viewpoints, URhGe\cite{AokiNature2001} and UCoGe\cite{HuyPRL2007} are located in special situations, not only in ferromagnetic (FM) superconductors, but also in all unconventional superconductors.
This is because the superconductivity is reentrant\cite{LevyScience2005} or is enhanced under a high magnetic field parallel to the $b$ axis ($H \parallel b$)\cite{AokiJPSJ2011}, and is extremely sensitive to the applied magnetic field. 
When the high field is tilted to only a couple of degrees to the $c$ axis, the reentrant or enhancement of superconductivity disappears abruptly\cite{LevyScience2005, AokiJPSJ2009}.
This strongly suggests that the SC pairing interaction would be field dependent; therefore, this feature is very useful for studying the SC pairing interaction, because it is precisely controllable.   
 
We studied the FM superconductor UCoGe, discovered by Huy {\it et al.} in 2007\cite{HuyPRL2007}, using nuclear magnetic resonance (NMR) and nuclear quadrupole resonance (NQR) measurements from a microscopic point of view\cite{OhtaJPSJ2010, HattoriJPSJ2014Rev}. 
In UCoGe, FM transition occurs at the Curie temperature $T_{\rm Curie} \simeq 3$ K with a small ordered moment $m_{0} \simeq 0.05 \mu_{B}$ pointing to the $c$ axis, and superconductivity sets in at $T_{\rm SC} \simeq 0.6$ K. 
From the detailed angle dependence of the nuclear spin-lattice relaxation rate ($1/T_1$) and AC magnetic-susceptibility measurements below 3 T, we found that magnetic fields along the $c$ axis ($H \parallel c$) strongly suppress the FM fluctuations with the Ising anisotropy along the $c$ axis and that superconductivity is observed in the limited magnetic-field region where the Ising-type FM fluctuations are active\cite{HattoriPRL2012}.
These experimental results suggest that the Ising-type FM fluctuations tuned by $H \parallel c$ induce spin-triplet superconductivity. 
In addition, we measured $1/T_1$ down to 2 K under the field along the $a$ and $b$ axes up to $\sim 11$ T, and reported that $T_{\rm Curie}$ is suppressed and $1/T_1$ at 2 K is enhanced in $H \parallel b$, although $T_{\rm Curie}$ and $1/T_1$ are unchanged in $H \parallel a$\cite{HattoriJPSJ2014}.
Therefore, we suggest that the enhancement of the superconductivity observed in $H \parallel b$ originates from the FM critical fluctuations induced by $H \parallel b$.

In the sister superconductor URhGe, the reentrant superconductivity (RSC) was observed at approximately 13 T in $H \parallel b$, where the magnetization along the $b$ axis shows a superlinear increase (metamagnetic transition)\cite{LevyScience2005}. 
Tokunaga {\it et al.} investigated the magnetic fluctuations around the RSC region from the measurement of $1/T_1$ and nuclear spin-spin relaxation rate $1/T_2$, and showed that FM critical fluctuations develop in the same limited field region where RSC is observed. 
In particular, by comparing the divergence in $1/T_1$ and $1/T_2$, they revealed that the longitudinal fluctuations along the applied $H$, which are detectable by $1/T_2$ measurements, are much more dominant than the transverse fluctuations detectable by $1/T_1$ and become critical in the RSC field region in URhGe\cite{TokunagaPRL2015}.
Therefore, it is crucial to measure $1/T_2$ in UCoGe to identify the dominant magnetic fluctuations in the $H$ region, where the superconductivity is enhanced by $H \parallel b$.
In addition, it is also important to determine the direction of the critical FM fluctuations in the high magnetic field region around and above the FM criticality.

In this paper, we report the $1/T_1$ and $1/T_2$ results obtained from the $^{59}$Co NMR measurements in the magnetic field $H$ up to 23 T and with the temperature $T$ down to 1.5 K, as well as the upper critical field $H_{\rm c2}$ of the same single-crystal sample with the AC susceptibility measurements under a magnetic field exactly parallel to the $a$ or $b$ axis.  
We found that $1/T_1$ and $1/T_2$ measured at 1.5  K show a clear peak at approximately 12.5 $ T$, where $T_{\rm Curie}$ is nearly suppressed to zero in $H \parallel b$, whereas $1/T_1$, $1/T_2$, and $T_{\rm Curie}$ are unchanged in $H \parallel a$.
From the analyses of the field dependence of $1/T_1$ and $1/T_2$, it is concluded that the Ising-type FM fluctuations along the $c$ axis become critical at around 12.5 T, in contrast to the case of URhGe. 
By using the field dependence of the critical behavior of $1/T_1T$, we compare the field dependence of the difference of $T_{\rm SC}$ along the $a$ and $b$ axes with a difference of $1/T_1(H)T$ at 1.5 K on the basis of the theoretical model.
It was discovered that the characteristic feature of the $H_{\rm c2}$ behaviors in UCoGe can be semi-quantitatively explained by the field dependence of $1/T_1T$, following the theoretical formalism with approximation. 
A similar clear relation between $T_{\rm SC}$ and $1/T_1T$ variation was observed in $H \parallel c$ as well as in the pressure-induced FM critical region.  
These results provide strong evidence that the superconductivity in UCoGe is induced by critical FM fluctuations.   

\section{Experimental Methods}
\subsection{Samples}
The present single-crystal sample was prepared by the Czochralski crystal pulling method in a tetra-arc furnace under high-purity argon, and was cut into cube-like shapes with dimensions $a \times b \times c = 1.20 \times 1.14 \times 1.20$ mm$^3$.
The electrical-resistivity measurement along the $b$ axis revealed a residual resistivity ratio RRR $\sim 98$ and $T_{\rm Curie} = 2.7$ K. The onset, midpoint, and zero-resistivity temperatures of the SC transition were 0.87 K, 0.74 K, and 0.64 K, respectively.    
Clear specific-heat jumps at $T_{\rm Curie}$ and $T_{\rm SC}$ and a large RRR attest to the high quality of the sample. 

\subsection{Measurement of AC susceptibility for the determination of $T_{\rm SC}(H)$}
We measured the AC susceptibility to determine $T_{\rm SC}(H)$ under a magnetic field ($H$). 
As reported previously\cite{AokiJPSJ2009}, $T_{\rm SC}(H)$ is extremely sensitive to the orientation of a single crystal against the applied $H$.
To align the crystal accurately and to measure the AC magnetic susceptibility down to 100 mK, the NMR coil with the crystal was mounted on a piezoelectric rotator (ANRv51/ULT/RES+) with an angular resolution of $\sim 0.1^{\circ}$, and all were  immersed into the $^3$He-$^4$He mixture to reduce the radio-frequency heating. 
As we can rotate over one axis only, we rotated the $ca (bc)$ plane for the $a (b)$ axis alignment to minimize the $c$-axis component.

\subsection{NMR measurements}
\begin{figure}[tbp]
\begin{center}
\includegraphics[width=8cm]{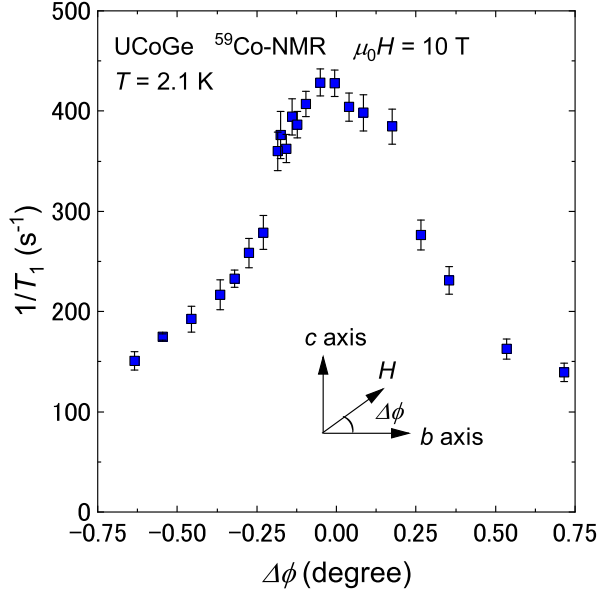}
\end{center}
\caption{Angle dependence of $1/T_1$ of $^{59}$Co in the $bc$ plane measured at 2.1 K for the determination of the $b$ axis. The peak angle was determined as the $b$-axis.   }
\label{fig1}
\end{figure}

As for the NMR measurement, the conventional spin-echo method was used.
The $^{59}$Co-NMR measurement down to 1.5 K was carried out up to 23 T. 
We used a 25 T cryogen-free SC magnet at the Institute for Materials Research at Tohoku University for NMR measurements above 15 T.
$1/T_1$ was measured by applying the saturation $\pi/2$ pulse at time $t$ prior to the $\pi/2 - \tau - \pi$ spin-echo sequence and by monitoring the recovery $R(t)$ of the spin-echo intensity as a function of $t$, and was evaluated by fitting $R(t)$ to the theoretical function\cite{NarathPR1967}.    
$1/T_2$ was measured by monitoring the decay of the spin-echo intensity $I(\tau)$ as a function of the interval time $\tau$ between the $\pi/2$ and $\pi$ pulses (Hahn method), and was evaluated by fitting $I(\tau)$ to the exponential function.
Satisfactory fitting by a single component of $T_1$($T_2$) was obtained in $R(t)$ ($I(\tau)$) in the entire $T$ and $H$ measurement region, indicating that magnetic fluctuations are homogeneous across the entire region of the sample.              
$1/T_1$ and $1/T_2$ are measured at the central transition peak of the $^{59}$ Co NMR spectrum in $H \parallel$ $a$ and $b$.  
The orientation of the external magnetic fields along the $a (b)$ axis in the normal state was carefully controlled {\it in situ} using a laboratory-built single-axis rotator by rotating the sample in the $ca (bc)$ plane. 
As reported previously\cite{HattoriJPSJ2014}, $1/T_1$ just below $T_{\rm Curie}$, which is governed by the FM critical fluctuation, is strongly suppressed when $H$ has a small $c$-axis component; therefore, we measured the angle dependence of $1/T_1$ in the $bc$ plane at 2.1 K and determined the $b$ axis from the peak of $1/T_1$, as shown in Fig.~\ref{fig1}.  

\section{EXPERIMENTAL RESULTS}

\subsection{AC susceptibility measurements in magnetic fields}

\begin{figure}[tbp]
\begin{center}
\includegraphics[width=8cm]{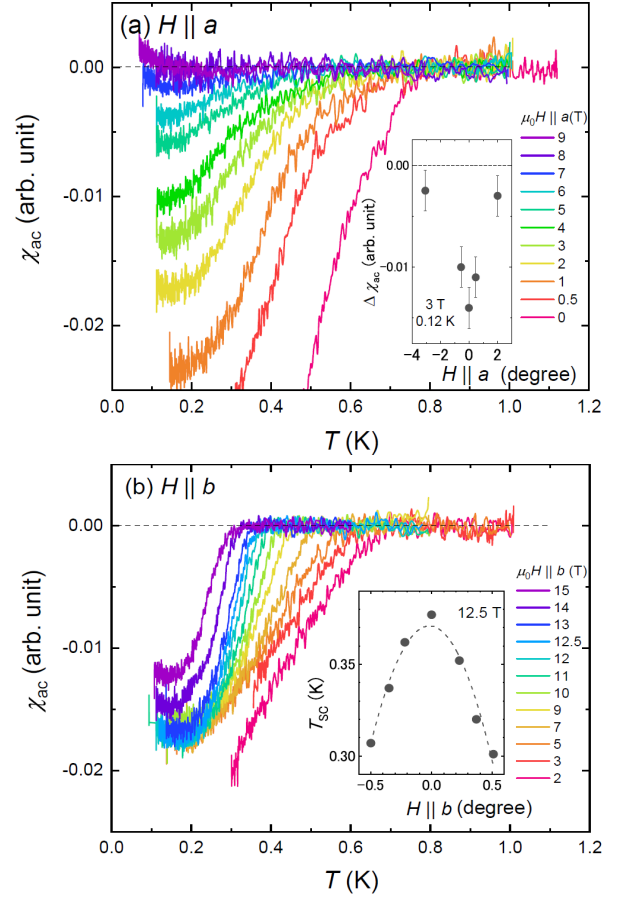}
\end{center}
\caption{The AC susceptibility measured at various fields in (a) $H \parallel a$ and (b) $H \parallel b$. The inset of (a) shows the angle dependence of the decrease in $\chi_{ac}$ below $T_{\rm SC}$ measured at 3 T and 0.12 K. The angle with the largest decrease was determined as the $a$-axis.  The inset of (b) shows the angle dependence of $T_{\rm SC}$ measured at 12.5 T. The angle with the highest $T_{\rm SC}$ was determined as the $b$ axis.  }
\label{fig2}
\end{figure}

\begin{figure}[tbp]
\begin{center}
\includegraphics[width=8cm]{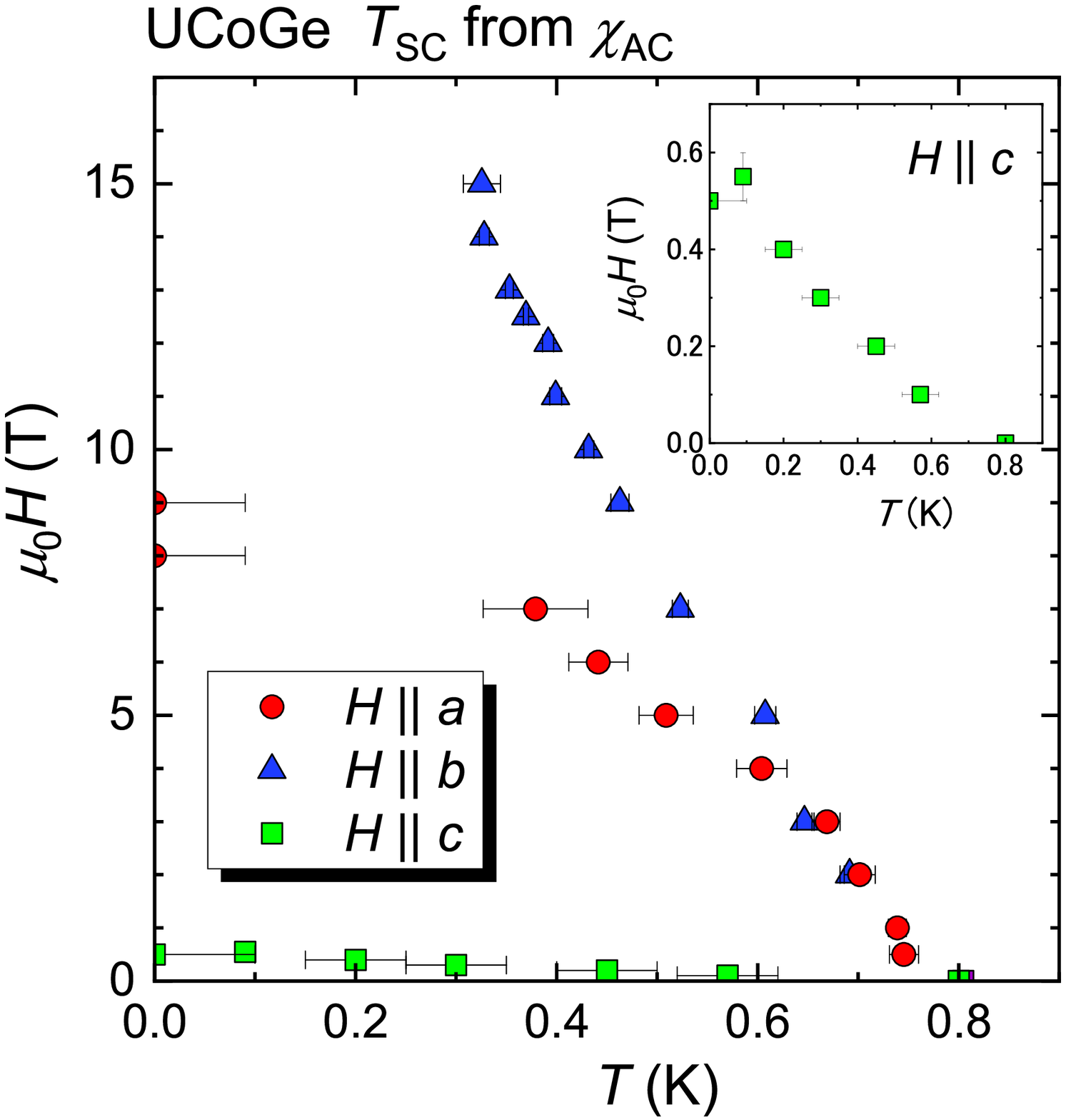}
\end{center}
\caption{Upper critical field $H_{\rm c2}$ along each crystal axis. $H_{\rm c2}$ was determined using the AC susceptibility measurements. The inset shows the zoom of $H_{c2}$ along the $c$ axis.    }
\label{fig3}
\end{figure}
   
Figures~\ref{fig2}(a) and (b) show the results of the AC magnetic-susceptibility measurements at various fields in (a) $H \parallel a$ and (b) $H \parallel b$. 
For the measurements, the $a$ axis was determined from the angle dependence of the decrease in $\chi_{\rm AC}$ below $T_{\rm SC}$ measured at 3 T, and the $b$ axis was determined from the angle dependence of $T_{\rm SC}$ measured at 12.5 T, as shown in the inset of each figure.  
Here, $T_{\rm SC}$ in each field was determined from the onset of the decrease in $\chi_{\rm AC}$ (Meissner signal) from the constant value. 
The experimental errors are determined from the noise level of $\chi_{\rm AC}$.  
Figure~\ref{fig3} shows the upper SC critical fields ($H_{\rm c2}$) in the present single-crystal sample of UCoGe determined with the above AC susceptibility measurement. 
The zero points in the $H$ axis with the experimental error of 0.09 K in Fig. \ref{fig3} mean that the Meissner signal was not observed down to 0.09 K in the $T$-scan measurement of $\chi_{\rm AC}$.
In addition,  $H_{\rm c2}$ value at $T \sim 0.09$ K in the inset was determined from the $H$-scan measurement of $\chi_{\rm AC}$.   

The $\mu_0 H_{\rm c2}$ shows a huge anisotropy: $\mu_0 H_{\rm c2}$ along the $b$ seems to be beyond 20 T, but $\mu_0 H_{\rm c2}$ along the $c$ is only 0.6 T, although the transport properties in UCoGe is rather three dimensional as reported previously\cite{HattoriPRL2012}.
The remarkable enhancement of $H_{\rm c2}$ characterized by the S-shaped $H_{\rm c2}$ behavior\cite{AokiJPSJ2009, WuNatComm2017} was not observed in the present sample. 
However, the different behavior in $H_{\rm c2}$ might be due to sample dependence and/or difference of the experimental methods. 
The comparison of $H_{\rm c2}$ values and the magnitude of the Meissner signals between $H \parallel a$ and $H \parallel b$ indicates that the superconductivity in the present sample also becomes robust in $H \parallel b$ as reported in the previous measurement\cite{AokiJPSJ2009, WuNatComm2017}.

\subsection{$1/T_1$ and $1/T_2$ in magnetic fields}
\begin{figure}[tbp]
\begin{center}
\includegraphics[width=8cm]{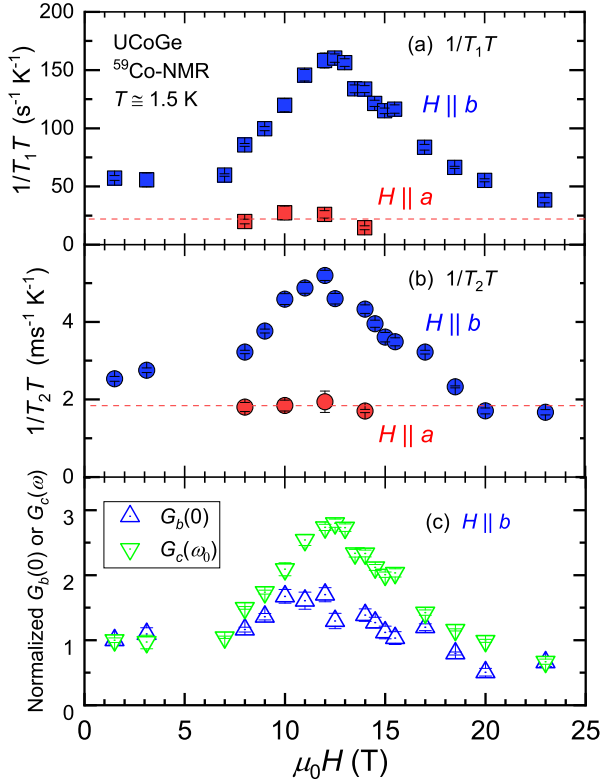}
\end{center}
\caption{Magnetic field ($H$) dependence of $1/T_1T$ (a) and $1/T_2T$ (b) measured in $H \parallel a$ and $H \parallel b$. (c) The $H$ dependence of $G_b(0)$ and $G_c(\omega_0)$ derived from $1/T_1T$ and $1/T_2T$ (see text).    }
\label{fig4}
\end{figure}
Figures \ref{fig4} (a) and (b) show the $H$ dependence of $1/T_1T$ and $1/T_2T$ at 1.5 K in $H \parallel a$ and $H \parallel b$ measured up to 23 T.     
As reported previously\cite{HattoriJPSJ2014}, $1/T_1T$ shows anisotropic behavior between the $a$ and $b$ axes.
It was clarified that both $1/T_1T$ and $1/T_2T$ in $H \parallel b$ show a maximum at around 12.5 T in a similar manner, although both are $H$ independent in $H \parallel a$.
When $1/T_2$ is determined by the electronic contribution, $1/T_1T$ and $1/T_2T$ measured in $H \parallel x$ are expressed as\cite{WalstedtPRL1967},
\begin{eqnarray*}
\lefteqn{\left(\frac{1}{T_1T}\right)_{x}} \\
& = & \frac{\gamma_n^2 k_{\rm B}}{(\gamma_e\hbar)^2} \sum_{{\bm q}} \left[
|A^{y}_{\rm hf}|^2\frac{\chi''_{y}({\bm q},\omega_0)}{\omega_0} + 
|A^{z}_{\rm hf}|^2\frac{\chi''_{z}({\bm q},\omega_0)}{\omega_0}\right] \\
& = & G_{\perp}(\omega_0), 
\label{eq:T1}
\end{eqnarray*} 
and
\begin{eqnarray*}
\lefteqn{\left(\frac{1}{T_2T}\right)_{x} =\frac{\gamma_n^2 k_{\rm B}}{(\gamma_e\hbar)^2} \lim_{\omega \rightarrow 0} \sum_{{\bm q}} \left[ |A^{x}_{\rm hf}|^2\frac{\chi''_{x}({\bm q},\omega)}{\omega}\right]} \\
& & +\left[I(I+1)-m(m+1)-1/2\right]\left(\frac{1}{T_1T}\right)_x\\
&=&G_{\parallel}(0)+\alpha G_{\perp}(\omega_0).
\end{eqnarray*}
Here, $\chi''_{i}({\bm q},\omega)$ ($i = x, y,$ and $z$) is the dynamical spin susceptibility components along the $i$ axis, and $\omega_0$ is the NMR frequency.
$G_{k}(\omega) = \int_{-\infty}^{\infty}\langle h_{k}(t) h_{k}(0)\rangle \exp{(i\omega t)}dt$ ($k = \parallel$ and $\perp$) is the spectral density of the fluctuating hyperfine field $h_k(t)$.
It should be noted that $1/T_1$ in $H \parallel x$ is determined by magnetic fluctuations perpendicular to $H$, but $1/T_2T$ in $H \parallel x$ can detect nearly static magnetic fluctuations parallel to $H$ in addition to an extra contribution from $1/T_1T$ in $H$.    
The coefficient of the second term $\alpha$ in the $1/T_2T$ is 15.5, for the central transition ($m_z = 1/2 \leftrightarrow 1/2$) of the Co NMR spectrum ($I$ =7/2).
When $H \parallel b$, $G_{\parallel}(0) = G_b(0)$ and $G_{\perp}(\omega_0) \sim G_{c}(\omega_0)$, because $\chi''_{c} \gg \chi''_{a}$ is shown from the axial dependence of $1/T_1$ at low $T$\cite{IharaPRL2010}. 
Using these formulae, the $H$ dependence of $G_{c}(\omega_0)$ and $G_{b}(0)$ normalized with each low-$H$ value is shown in Fig. \ref{fig4} (c).  
The $H$ dependence of $G_{c}(\omega_0)$ exhibits a clear maximum around FM criticality, although $G_{b}(0)$ shows a small hump. 
It is obvious that the critical fluctuations originate from $G_{c}(\omega_0)$; the Ising FM magnetic fluctuations along the $c$ axis, inherent to UCoGe, become critical for the application of $H \parallel b$.
This situation is different from the FM criticality in URhGe induced by metamagnetism in $H \parallel b$, where the magnetic fluctuations along the $H$ direction become critical.
The difference is related to the anisotropy and the character of the magnetization in two compounds; the Ising anisotropy along the $c$ axis is much stronger in UCoGe. 
This is shown from the absence of the metamagnetic transition up to 40 T in UCoGe, and the metamagnetic transition was observed at 12 T in URhGe\cite{KnafoPRB2012}.
The different magnetic properties between UCoGe and URhGe becomes more transparent from the recent study on UCo$_{1-x}$Rh$_x$Ge, although the 5$f$ electrons are in the itinerant regime in both compounds\cite{PospisilPRB2020}.     

\begin{figure}[tbp]
\begin{center}
\includegraphics[width=8.5cm]{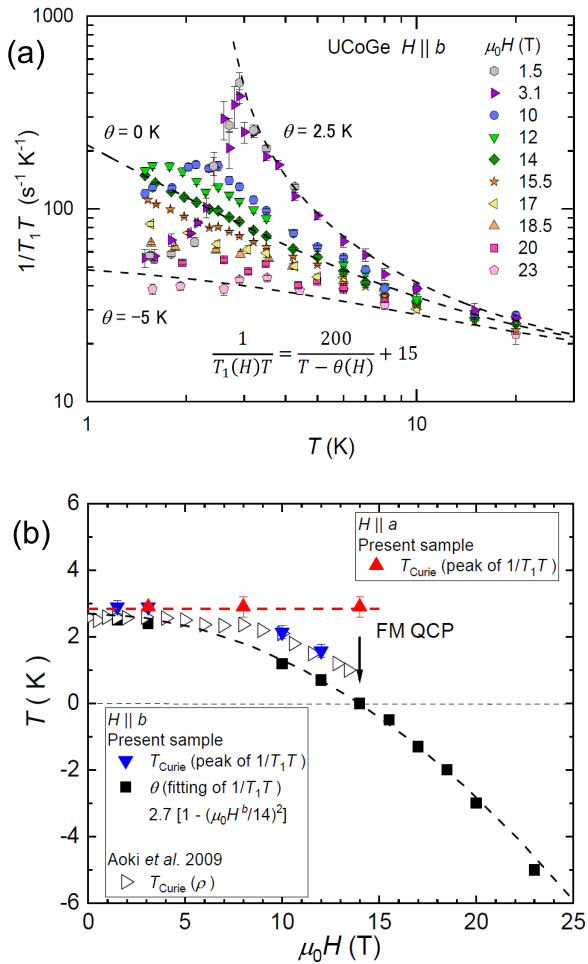}
\end{center}
\caption{(a) Temperature dependence of $1/T_1T$ of $^{59}$Co measured at various magnetic fields from $\mu_0 H$ = 1.5 to 23 T in $H \parallel b$. The phenomenological fitting of the experimental $1/T_1T$ is shown by the dotted curves with $\theta$ = 2.5, 0, and -5 K. (b) $H$ dependence of $T_{\rm Curie}$ determined by the peak or maximum of $1/T_1T$ measured in $H \parallel a$ and $H \parallel b$ in the present sample. The characteristic temperature $\theta(H)$ corresponding to the magnetic anomaly in the present sample is also plotted.  The $T_{\rm Curie}$ determined by the bending of the electric resistance is also plotted for comparison.      }
\label{fig5}
\end{figure}
Next, we measured the temperature ($T$) dependence of $1/T_1T$ in various magnetic fields along the $b$ axis to investigate the critical behavior of the Ising FM fluctuations along the $c$ axis.           
Figure \ref{fig5}(a) shows the $T$ dependence of $1/T_1T$ in various magnetic fields along the $b$ axis.
$1/T_1T$ shows a clear peak at 1.5 T and 3 T and a broad maximum at 10 and 12 T at $T_{\rm Curie}$.
In the $H$ region greater than 18.5 T, $1/T_1T$ becomes almost constant below 3 K, and the constant values of $1/T_1T$ decrease with increasing $H$.    
It is noteworthy that the $T$ dependence of $1/T_1T$ in the entire $H$ region is well fitted by the following equation:      
\begin{equation*}
\left(\frac{1}{T_1(H)T}\right)_{H \parallel b}   =  \frac{200}{T- \theta(H)} + 15 
\label{eq:(3)}
\end{equation*} 
only by changing $\theta(H)$. 
Here, $\theta(H)$ is a parameter of how close the system is to the FM criticality.
The evaluated $\theta(H)$ from the $T$ dependence of $1/T_1(H)T$ is plotted in Fig.~\ref{fig5} (b), along with the Curie temperature $T_{\rm Curie}$ determined by the peak of $1/T_1T$.
It is noted that the evaluated $\theta(H)$ systematically changes from positive to negative values and becomes zero at approximately 14 T, where $T_{\rm Curie}$ is assumed to be zero.
This shows the presence of the FM critical point at approximately 14 T, and that the $1/T_1(H)T$ can trace the FM criticality systematically.
In fact, the $H$ dependence of $\theta$ can be well fitted with the $H$ quadratic behavior, which is expected in the mean-field theory\cite{MineevPRB2015}.     
In the self-consistent renormalization theory describing weak ferromagnetism, $1/T_1T$ is proportional to the static susceptibility $\chi$ in a three-dimensional system\cite{TMoriya1991}. Therefore, it is experimentally indicated that $1/T_1T$ in $ H \parallel b$ is a good measure of magnetic susceptibility $\chi_c$ related to the FM criticality.         

\section{DISCUSSION}
In this section, we present a semi-quantitative discussion of the relationship between FM critical fluctuations and superconductivity.
As mentioned above, we have shown the relationship between the Ising-type FM fluctuations and the superconductivity in UCoGe based on the $^{59}$Co NMR measurements.
From a theoretical perspective, a theoretical models of spin-triplet superconductivity observed in U-based ferromagnets were discussed by Mineev \cite{MineevPRB2011, MineevPUsr2017, MineevPRB2021}, and Tada and Fujimoto\cite{HattoriPRL2012, TadaJPCS2013, TadaPRB2016}, Hattori and Tsunetsugu\cite{HattoriPRB2013URhGe}, independently.
Mineev discussed the triplet pairing triggered by the exchange of magnetic fluctuations with a phenomenological fluctuation spectrum\cite{MineevPRB2011, MineevPUsr2017, MineevPRB2021}. 
Tada and Fujimoto showed that the experimentally observed anomalous $H_{c2}$ behaviors against $T$ and the angle in the $bc$ plane are well reproduced by the theoretical model with appropriately taking into account the critical FM fluctuations observed in experiments\cite{HattoriPRL2012, TadaJPCS2013, TadaPRB2016}.      
Hattori and Tsunetsugu discussed the RSC in URhGe by strong attractive interactions generated by soft magnons in the Ising systems with transverse fields\cite{HattoriPRB2013URhGe, SuzukiJPSJ2020}.  
These models were developed to understand the unusual response of superconductivity against the applied $H$ observed in U-based FM superconductors.

\begin{figure}[tbp]
\begin{center}
\includegraphics[width=8.5cm]{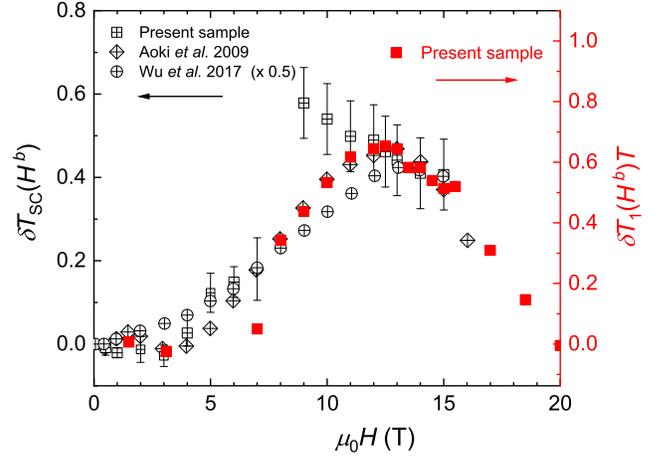}
\end{center}
\caption{The $H$ dependence of the difference of $T_{\rm SC}$ between $H \parallel b$ and $H \parallel a$ normalized by zero-field $T_{\rm SC}(0)$, $[T_{\rm SC}(H^b) - T_{\rm SC}(H^a)]/T_{\rm SC}(0)  \equiv \delta T_{\rm SC}(H^b)$ on the present sample and the $H$ dependence of the normalized difference of $T_{\rm SC}$ evaluated from the previous reports\cite{AokiJPSJ2009, WuNatComm2017} are also plotted for the left axis, and the vertical scale of the results reported by Wu {\it et al.} is reduced to half for the comparison. The $H$ dependence of $[T_1(H \sim 0)T - T_1(H^b)T] / T_1(H \sim 0)T \equiv \delta T_1(H^b)T$ measured at 1.5 K on the present sample is plotted on the right axis. }
\label{fig6}
\end{figure}

When an orbital suppression of superconductivity is neglected and the largest critical temperature corresponds to the pairing of quasiparticles with spin up-up and spin down-down, Mineev expressed BCS-type $T_{\rm SC}$ formula in a weak coupling interaction as \cite{MineevPUsr2017, MineevPRB2021},
\begin{equation*}
T_{\rm SC}(H) = \varepsilon \exp{\left(-\frac{1}{g(H)}\right)},
\end{equation*}
where $g(H)$ is the pairing interaction for the single-band model, in which only spin up-up pairing is formed, and is considered to be a function of $H$, and $\varepsilon$ is the energy cutoff of the pairing interaction.
In $H \parallel a (b)$, $g$ is expressed as $g \sim [\langle N_0(\mbox{\boldmath$k$})\chi_c^u(H^{a(b)}) \rangle \cos^2{\varphi}+\langle N_0(\mbox{\boldmath$k$})\chi_{a(b)}^u(H^{a(b)}) \rangle \sin^2{\varphi}]I^2$, where $I$ is the coupling constant of electrons with spin fluctuations, $\tan{\varphi} = H^{a(b)}/h$, and $h$ is the exchange field acting on the electron spins. 
The angular brackets denote the average over the Fermi surface, and $N_0(\mbox{\boldmath$k$})$ is the angular dependent density of the electronic states on the Fermi surface. 
$\chi_c^u$ and $\chi_{a(b)}^u$ are odd in the momentum part of the static susceptibilities along the $c$ and $a$ ($b$) axes, respectively. 
In UCoGe near $T_{\rm SC}$, the Ising anisotropy in the static susceptibility is so large that $\chi_{a(b)}^u$ is negligibly smaller than $\chi_c^u$, and $\varphi \ll 1$ such that $\cos{\varphi} \approx 1$,
since the exchange field $h$ is considered to be $\sim 50$ T, much larger than $H^{a(b)}$, from the metamagnetic behavior in $M /H$ in $H \parallel b$\cite{KnafoPRB2012}. 
Thus, we approximately obtain        
\begin{equation*}
T_{\rm SC}(H^{a(b)}) \sim \varepsilon \exp{\left(-\frac{\alpha_1}{\chi_c(H^{a(b)})}\right)}.
\end{equation*}
Here, $\alpha_1$ is a constant, and we assume that $\chi_c$ is proportional to $1/T_1T$ from the $T$ and $H$ dependence of  $1/T_1T$, as shown above. 
If we take the difference between $T_{\rm SC}(H^b)$ and $T_{\rm SC}(H^a)$, the orbital suppression can be canceled out in the lower $H$ region, and the superconductivity enhanced by the FM criticality can be mainly extracted, although the cancelation of the orbital suppression is insufficient in the higher $H$ region.
Therefore, $T_{\rm SC}(H^b) - T_{\rm SC}(H^a)$ normalized by $T_{\rm SC}(0)$, which is defined as $\delta T_{\rm SC}(H^b)$, is extracted and can be expressed in the lowest order as 
\begin{eqnarray*}
\frac{T_{\rm SC}(H^b) -T_{\rm SC}(H^a)}{T_{\rm SC}(0)} &\sim& \frac{\exp{\left(-\frac{\alpha_1}{\chi_c(H^b)}\right)}-\exp{\left(-\frac{\alpha_1}{\chi_c(H^a)}\right)}}{\exp{\left(-\frac{\alpha_1}{\chi_c(0)}\right)}}\\
\propto \alpha_1 \left[ \frac{1}{\chi_c(0)}-\frac{1}{\chi_c(H^b)} \right]  &\propto& \frac{T_1(0)T - T_1(H^b)T}{T_1(0)T}, 
\end{eqnarray*}
by taking $\chi_c(H^a) \sim \chi_c(0)$ on the basis of $H^a$ independent $1/T_1T$, and using the experimental relation of  $\chi_c(H^b) \propto (T_1(H^b)T)^{-1}$. 
Figure \ref{fig6} shows the $H$ dependence of the $\delta T_{\rm SC}(H^b) \equiv [T_{\rm SC}(H^b) -T_{\rm SC}(H^a)]/T_{\rm SC}(0)$ in UCoGe reported so far for the left axis\cite{AokiJPSJ2011, WuNatComm2017} and that of the normalized $T_1T$ difference, $\delta T_1(H^b)T \equiv [T_1(0)T - T_1(H^b)T]/T_1(0)T$ at 1.5 K measured on the present UCoGe [Fig.~\ref{fig4}(a)] for the right axis.     
It is interesting that $\delta T_{\rm SC}(H^b)$ and $\delta T_1(H^b)T$ show almost the same $H$ dependence, as suggested by the theoretical model\cite{MineevPUsr2017, MineevPRB2021}.

\begin{figure}[tbp]
\begin{center}
\includegraphics[width=8.5cm]{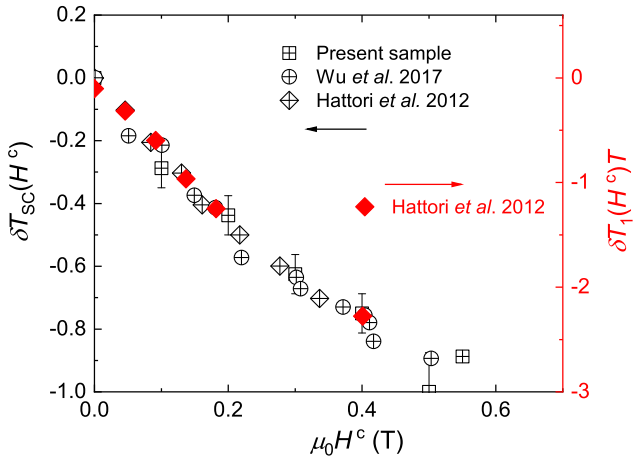}
\end{center}
\caption{The $H$ dependence of $[T_{\rm SC}(H^c) - T_{\rm SC}(0)]/T_{\rm SC}(0)  \equiv \delta T_{\rm SC}(H^c)$ on the present sample and its $H$ dependence evaluated from previous reports are also plotted for the left axis\cite{HattoriPRL2012,WuNatComm2017}. The $H$ dependence of $[T_1(H \sim 0)T - T_1(H^c)T] / T_1(H \sim 0)T \equiv \delta T_1(H^c)T $ measured at 1.5 ~ K reported by our group is plotted on the right axis\cite{HattoriPRL2012}. The $H^c$ dependence of $1/T_1T$ is evaluated from the angle dependence of $1/T_1T$ in the $bc$ plane.  }
\label{fig7}
\end{figure}

We apply this theoretical model to other two cases observed in UCoGe.
One is the steep decrease of $H_{\rm c2}$ in $H \parallel c$. 
It is obvious that such a steep decrease cannot be explained only by the orbital suppression effect, which was experimentally demonstrated by Wu {\it et al. }\cite{WuNatComm2017}, and we have shown the relationship between the suppression of the Ising fluctuations and the steep decrease of $T_{\rm SC}$ in $H \parallel c$, as pointed out previously\cite{HattoriPRL2012}.
Following the above theoretical analysis, $T_{\rm SC}(H^c)$ in $H \parallel c $ is approximately expressed as\cite{MineevPUsr2017, MineevPRB2021}:
\begin{eqnarray*}
\lefteqn{T_{\rm SC}(H^{c})} \\
&=& \varepsilon \exp{\left(-\frac{1}{\langle N_0(\mbox{\boldmath $k$})\chi_{c}^u(H^{c}) \rangle I^2}\right)} \sim \varepsilon \exp{\left(-\frac{\alpha_2}{\chi_c(H^{c})}\right)}  .
\end{eqnarray*}
Here, we assume that UCoGe is still close to the FM criticality, where the magnetic coherence length $\xi_m$ is much larger than the lattice constant ($\xi k_{\rm F} \gg 1$).  
Using this relation and $\chi_c(H^c) \propto (T_1(H^c)T)^{-1}$, the normalized decreasing rate $T_{\rm SC}$ against $H^c$, defined as $\delta T_{\rm SC}(H^c)$ is expressed as
\begin{eqnarray*} 
\delta T_{\rm SC}(H^c) & \equiv &\frac{T_{\rm SC}(H^c) -T_{\rm SC}(0)}{T_{\rm SC}(0)} \\
 &\propto& \frac{T_1(0)T - T_1(H^c)T}{T_1(0)T} \equiv \delta T_1(H^c)T. 
\end{eqnarray*}
Figure \ref{fig7} shows the $H^c$ dependence of $\delta T_{\rm SC}(H^c)$ in UCoGe reported so far for the left axis\cite{HattoriPRL2012, WuNatComm2017} and that of $\delta T_1(H^c)T$ at 1.5 K reported previously for the right axis\cite{HattoriPRL2012}.  
The $H^c$ dependence of $T_1T$ was measured in $H \parallel b$ by adding a small $H^c$. 
This corresponds to the $T_1T$ measurements by the rotating $H$ from the $b$ to $c$ axis, and the details are reported in \cite{HattoriPRL2012}.

\begin{figure}[tbp]
\begin{center}
\includegraphics[width=8.5cm]{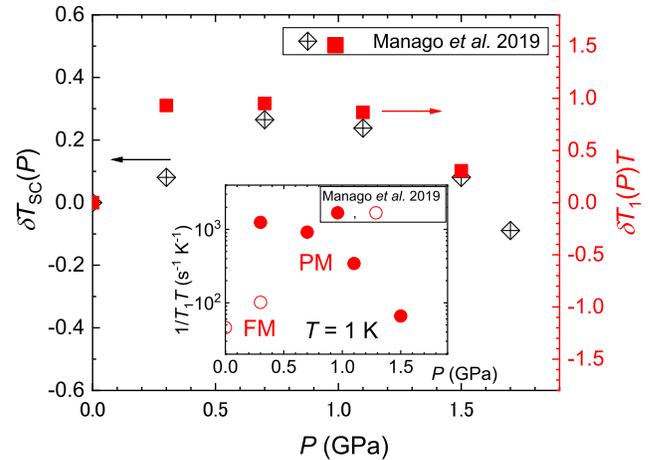}
\end{center}
\caption{The pressure ($P$) dependence of $[T_{\rm SC}(P) - T_{\rm SC}(0)]/T_{\rm SC}(0) \equiv \delta T_{\rm SC}(P) $ in the zero field is plotted on the left axis\cite{ManagoPRB2019P, ManagoJPSJ2019}. The $P$ dependence of $[T_1(0)T - T_1(P)T] / T_1(0)T \equiv \delta T_1(P)T$ measured at 1 K on the same sample of the $P$ dependence of $T_{\rm SC}$ is plotted on the right axis\cite{ManagoPRB2019P, ManagoJPSJ2019}. The inset shows the $P$ dependence of $1/T_1T$ measured at 1 K. PM (FM) means $1/T_1T$ in the paramagnetic (ferromagnetic) state.   }
\label{fig8}
\end{figure}

The other case is the pressure ($P$) induced FM criticality.
We studied the relationship between $P$-induced FM criticality and superconductivity by $^{59}$Co-NQR measurements in a zero magnetic field\cite{ManagoPRB2019P, ManagoJPSJ2019}.
The superconductivity is also enhanced by the $P$-induced FM critical region, which has been reported previously\cite{ HassingerJPSJ2008, SlootenPRL2009, BastienPRB2016}.
This suggests that the pairing interaction can be tuned by pressure, as in the case of $H \parallel b$.
We assume that $T_{\rm SC}(P)$ in $H = 0$ can be approximated as:
\begin{equation*}
T_{\rm SC}(P) \sim \varepsilon \exp{\left(-\frac{\alpha_3}{\chi_c(P)}\right)}.
\end{equation*} 
as in the case of $H \parallel b$.  
Following the above analysis, the normalized $T_{\rm SC}$ variation by $P$, defined as $\delta T_{\rm SC}(P)$ can be expressed as
\begin{eqnarray*}
\delta T_{\rm SC}(P)  & \equiv & \frac{T_{\rm SC}(P) -T_{\rm SC}(0)}{T_{\rm SC}(0) } \\
                              &\propto& \frac{T_1(0)T - T_1(P)T}{T_1(0)T} \equiv \delta T_1(P)T.
\end{eqnarray*}              
In Fig.~\ref{fig8}, the pressure dependence of $\delta T_{\rm SC}(P)$ evaluated from our UCoGe results is plotted on the left axis, and that of $\delta T_1(P)T$ measured at 1 K is plotted on the right axis\cite{ManagoPRB2019P, ManagoJPSJ2019}.
It should be noted that the $1/T_1T$ measured by NQR is determined by the $c$-axis fluctuations, because the principal axis of the electric quadrupole interaction is almost parallel to the $a$ axis.
As the FM and paramagnetic (PM) signals coexist at $P$ = 0.3 ~ GPa, we take an average for $T_1(0.3)T$.    
It is noteworthy that the good relationship between the variation in $\delta T_{\rm SC}(P)$ and that in $\delta T_1(P)T$ holds beyond the FM criticality.
This indicates that the FM fluctuations are intrinsically important, and that the presence of the ordered moments is not very sensitive to the superconductivity in UCoGe.

It is quite interesting that a good relation holds between $\delta T_{\rm SC}$ and $\delta T_1T$ in all three cases, where FM fluctuations are tuned by different external perturbations.
Since $1/T_1T$ probes the low-energy spin fluctuations, it was experimentally shown that this SC mechanism becomes dominant in UCoGe near the magnetic instability, where the low-magnetic fluctuations are enhanced. 
The SC mechanism by the electron-electron interactions, other than the electron-phonon interactions, was first pointed out by Kohn and Luttinger\cite{KohnPRL1965}. 
They pointed out that $T_{\rm SC}$ by their mechanism was so low in metals ($T_{\rm SC}$ was estimated as only $\sim 1$ mK in $s$-electron metals) but pointed out the possibility of ``high'' $T_{\rm SC}$ in metals where a flattening of the Fermi surface or an abnormally high density of states could be realized. 
This means that their mechanism becomes more effective in metallic compounds near the magnetic instability, where magnetic susceptibility and density of states become enhanced. 
It is reasonable to consider that the magnetic compounds, in which critical behaviors in $1/T_1$ are observed, show superconductivity also from their theoretical point of view.

It is also noted that the ratio between the left and right axes is the same as that in Fig.~\ref{fig6} (b) and (c) ($\sim 3$), and the ratio in Fig.~\ref{fig6} (a) is $\sim 1.4$, indicating that the proportional constants are nearly the same in these three cases.   
The good relation between $T_{\rm SC}$ and $T_1T$ variations strongly suggests that $T_{\rm SC}$ is determined by the FM fluctuations and is approximately expressed as $T_{\rm SC} \sim \varepsilon \exp{\left(-1/g \right)}$, where $g$ is intimately related to the anisotropic Ising spin susceptibility along the $c$ axis $\chi_c$ in UCoGe.

It is expected that the character of the spin susceptibility is highly dependent on SC compounds, even in U-based FM superconductors.
Although all U-based FM superconductors are considered as itinerant $f$-electron metals\cite{ShickPRB2002, ShickPRB2004, TaupinPRB2015, WilhelmPRB2017, FujimoriPRB2015}, the itinerant-localized components are different within them.   
Since both the FM and SC transitions occur far below the Kondo coherent temperature and the ordered moment in the FM state is small (0.05 $\mu_B$), the itinerant character is dominant in UCoGe.   
In contrast, the FM ordered moments in URhGe and UGe$_2$ are nearly one order of magnitude larger than those in UCoGe, thus the localized character is stronger in URhGe, and UGe$_2$.
This is the reason why a clear metamagnetic transition was observed in these two compounds\cite{AokiJPSJ2019Rev}.
It is important to check whether similar semi-quantitative analyses hold for URhGe and UGe$_2$. 

Finally, we comment on the advantage of the study of FM superconductors.
As the possibility of the spin-fluctuation-mediated superconductivity was pointed out\cite{MiyakePRB1986, ScalapinoPRB1986, PinesPhysicaB1990, MoriyaJPSJ1990}, a lot of studies have been done to clarify the relationship between the spin fluctuation and superconductivity, and in most of cases the relationship between antiferromagnetic (AFM) fluctuations and superconductivity has been investigated mainly on Ce-based heavy-ferimon (HF), cuprates, and Fe-based superconductors\cite{MoriyaRPP2003, PinesJPCB2013,NakaiPRB2013}. 
Although the superconductivity in these systems can be enhanced by critical fluctuations around the AFM critical points, the spin fluctuations cannot be tuned by $H$ because the AFM fluctuations are not coupled with $H$. 
This point is crucially different from the present case of the FM superconductors. 
We point out that FM superconductors are a key system for understanding the mechanism of spin-fluctuation-mediated superconductivity.  
It is also important to note that the pairing interactions originate from the spin-spin exchange process between the nearly localized 5$f$ electrons through the conduction electrons\cite{MineevPUsr2017}.
This process is the same as the Ruderman-Kittel-Kasuya-Yosida interaction\cite{RudermanPhysRev1954, YosidaPhysRev1957}, which also plays an important role in the realization of the HF state.
We point out that this is the reason why spin-fluctuation-mediated superconductivity has often been observed in HF compounds.      
It is also noted that the pairing interaction is cut off by the band energy of itinerant quasi-particles in the theoretical model\cite{MineevPUsr2017, MineevPRB2021}. 
We consider this to be the reason why $T_{\rm SC}$ of most HF superconductors has a linear relationship with their Fermi energy ($T_{\rm F}$), which is known as the Uemura plot\cite{UemuraPhysicaC1991}, the empirical relation holding in various unconventional superconductors.

\section{CONCLUSION}
We measured $1/T_1$ and $1/T_2$ of $^{59}$Co in $H \parallel a$ and $H \parallel b$ up to 23 T, as well as $H_{\rm c2}$ in each crystalline axis under the precise alignment of the single-crystal UCoGe.
Although the $1/T_1T$ and $1/T_2T$ at 1.5 K are almost $H$ independent in $H \parallel a$, it was found that the FM criticality in both $1/T_1T$ and $1/T_2T$ was observed in $H \parallel b$. 
From the comparison between $1/T_1T$ and $1/T_2T$, it can be concluded that the FM criticality arises from the Ising spin susceptibility along the $c$ axis $\chi_c$, and that $1/T_1T$ is a good measure of $H$ and $T$ dependence of $\chi_c$.
Based on the developed weak-coupling theory\cite{MineevPUsr2017, MineevPRB2021} and using the phenomenological relation $(T_1T)^{-1} \propto \chi_c$, we examined the relationship between $T_{\rm SC}$ and FM fluctuations derived from $1/T_1T$ when the FM fluctuations are enhanced by the suppression of $T_{\rm Curie}$ in $H \parallel b$, and are suppressed by $H \parallel c$. 
It was revealed that the variations of $T_{\rm SC}$ and FM fluctuations are well understood by this theoretical model at a semi-quantitative level.
A similar good relationship was also confirmed when the FM criticality was induced by pressure $P$.
The present semi-quantitative discussion concludes that superconductivity in UCoGe is induced by FM fluctuations tunable by $H$ and $P$.

\begin{acknowledgments}
One of the authors (K. I. ) wish to acknowledge V. P. Mineev for the theoretical inputs and fruitful discussions, and K. Deguchi, T. Yamamura, and N. K. Sato for the long-term collaboration on UCoGe.
The authors thank Y. Ihara, Y. Tokunaga, S. Yonezawa, Y. Maeno, H. Harima, H. Ikeda, S. Fujimoto, Y. Tada, A. de Visser, J-P. Brison, G. Knebel, and J. Flouquet for fruitful discussions. 
This work was supported by the Kyoto University LTM Center, Tohoku University (Project No. 202012-HMKPB-0052, 202012-IPKAC-0008), Grants-in-Aid for Scientific Research (Grant No. JP15H05745, JP17K14339, JP19K03726, JP16KK0106, JP19K14657, JP19H04696, JP20H00130, and JP20KK0061, and a Grant-in-Aid for JSPS Research Fellows (Grant No. JP20J11939) from JSPS.
We would like to thank Editage (www.editage.com) for English language editing.
\end{acknowledgments}

%

\end{document}